\documentclass{elsart}

\usepackage{graphicx}

\begin{document}

\begin{frontmatter}

\title{Cross-talk suppressed multi-anode MCP-PMT}

\author[Nagoya]{K.~Inami},
\author[Nagoya]{T.~Mori},
\author[Nagoya]{T.~Matsumura},
\author[Nagoya]{K.~Kurimoto},
\author[Nagoya]{S.~Hasegawa},
\author[Nagoya]{Y.~Suzuki},
\author[Nagoya]{T.~Murase},
\author[Nagoya]{Y.~Yurikusa},
\author[Nagoya]{M.~Akatsu},
\author[Nagoya]{Y.~Enari},
\author[Nagoya]{T.~Hokuue},
\author[Nagoya]{A.~Tomita},
\author[Nagoya]{N.~Kishimoto},
\author[Nagoya]{T.~Ohshima},
\author[HPK]{T.~Ihara},
\author[HPK]{H.~Nishizawa}
\address[Nagoya]
{Department of Physics, Nagoya university, Chikusa, Nagoya 464-8602, Japan}
\address[HPK]{Electron Tube Division, Hamamatsu Photonics K.K., 
314-5 Shimokanzo, Iwata, Shizuoka 438-0193, Japan}

\begin{abstract}
We have developed a 4-channel multi-anode MCP-PMT, SL10, which exhibits
a performance of $\sigma_{\rm TTS} \simeq 30$ ps for single photons with
$G \sim 10^6$ and $QE=20\%$ under a magnetic field of B $\leq 1.5 T$. 
The cross-talk among anodes has been extensively studied. We have taken
two measures to suppress it: one is to configure the SL10 to an 
effectively independent 4 small pieces of MCP-PMT's by segmenting 
an electrode of the second MCP-layer; the other is to 
use a constant fractional discriminator. Remarkable improvement has been
achieved. 
\end{abstract}

\begin{keyword}
MCP-PMT \sep cross-talk \sep TOP counter
\end{keyword}
\end{frontmatter}

\section{Introduction}

A micro-channel plate (MCP) photo-multiplier tube (PMT) provides
 a good time response. Our R\&D work had obtained a transit time spread
(TTS) of $\sigma_{\rm TTS}=30-35$ ps for single photons with a channel-hole
size of $<$ 10 $\mu$m$^{\phi}$ without a magnetic field (B) or with 
B$\leq$1.5 T strength~\cite{MCPtiming}. Such an MCP-PMT made it possible 
to realize $\sigma_{\rm TOF}=5$ ps of a TOF counter for 4 GeV/c pion 
beams~\cite{5psTOF}. 
Furthermore, an MCP-PMT with a multi-alkali photocathode maintains its timing
performance up to an integrated irradiation of $\simeq 2.8\times 10^{14}$
photons/cm$^2$ or an integrated output charge amount of 
$Q\simeq 2.6$ C/cm$^2$ under an irradiation 
rate of $4 \sim 20 \times 10^4$ photons/cm$^2$/s~\cite{MCPlifetime}. 
The MCP-PMT having these superb properties lets us develop
a new type of Cherenkov ring imaging counter, 
a time-of-propagation (TOP) counter~\cite{top},
for particle identification at a planned Super-KEKB factory with an 
expected luminosity of $\sim 10^{36}$ /cm$^2/$sec, 
which is 50-times as intense as the achieved luminosity of \
the current KEKB factory, 
and is expected to survive for more than 14 years under this situation. 

We here report on R\&D results of a newly developed MCP-PMT,
which is equipped with 4 anodes rather than a single anode, and 
shaped to be square, not round, from the required performance
on the TOP counter. 
When setting a multi-anode structure, an additional R\&D issue arises: 
it shows a cross-talk effect. 
While cross-talk basically occurs among the anodes, 
its influence is in most cases is harmless for measurements of 
the incident-particle energies
or a number of tracks in terms of the pulse-height of output signals. 
However, for a precise time measurement, as discussed here, 
the cross-talk deteriorates the timing performance, 
and sometimes makes it ineffectual, in much the same way that occurs for 
a vacuum PMT, as reported~\cite{X-PMT}. 
Based on our tests for various modified versions of the MCP-PMT, 
we take a specific measure to 
reduce the cross-talk while keeping its time resolution, as reported below. 
(We hereafter abbreviate MCP-PMT as simply PMT.)

\section{Multi-anode MCP-PMT: SL10}

\begin{figure}[h]
\centerline{
\includegraphics[width=5cm]{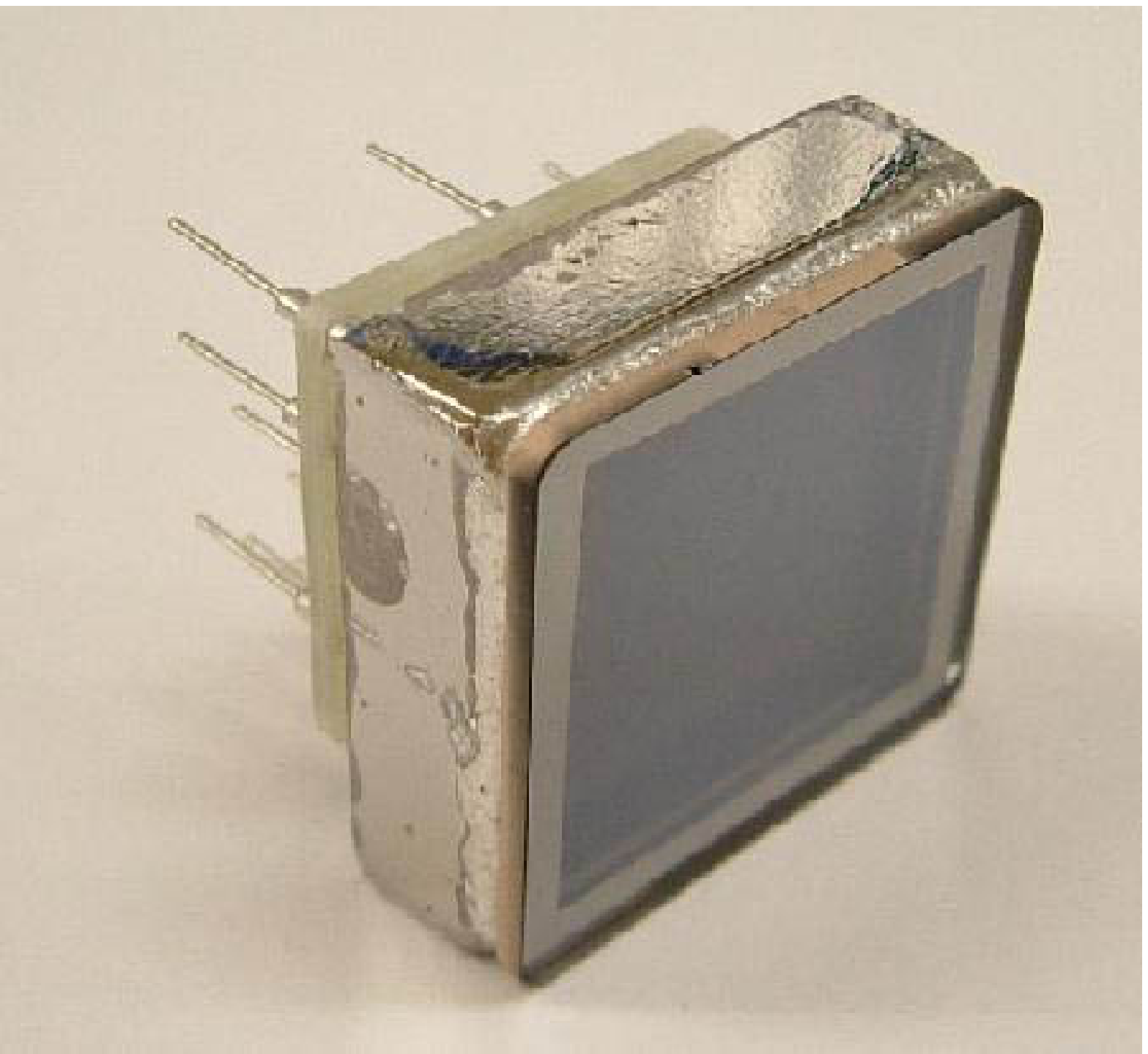}~~~~
\includegraphics[width=3.5cm]{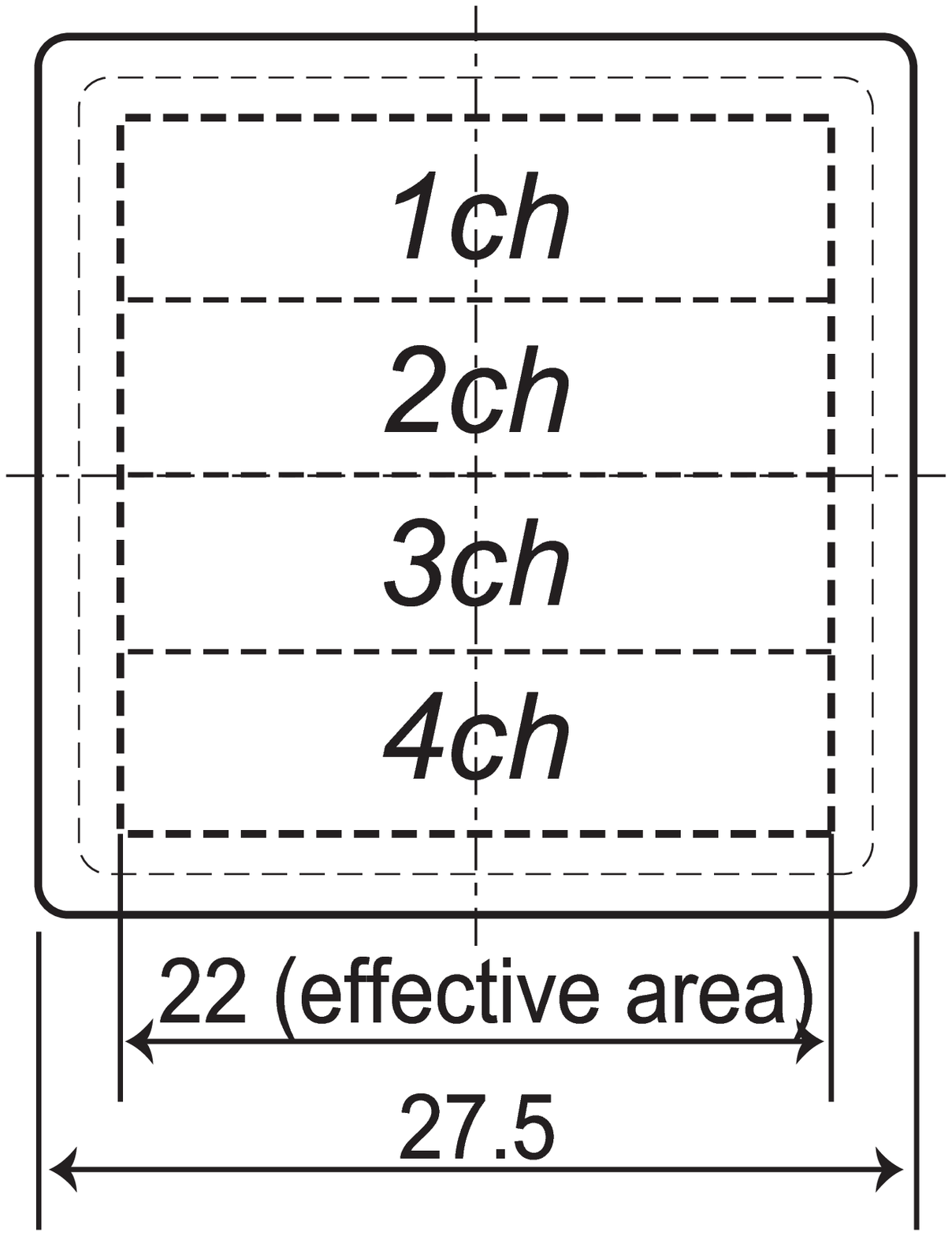}
}
\caption{\small Picture of SL10 and the anode layout.}
\label{F-1}
\end{figure}

We manufactured two versions of 4-channel linear-array multi-anode MCP-PMT's, 
and named them SL10, as can be seen in Fig.~\ref{F-1}. 
The difference in the versions only concerns the electrode of 
the second MCP-layer, facing to the anodes: 
SL10-s has a uniform electrode over the whole MCP surface, while SL10-m has 
an electrode subdivided into 4 segments facing the individual 4 anodes. 

\begin{figure}[h]
\centerline{
\includegraphics[width=7cm]{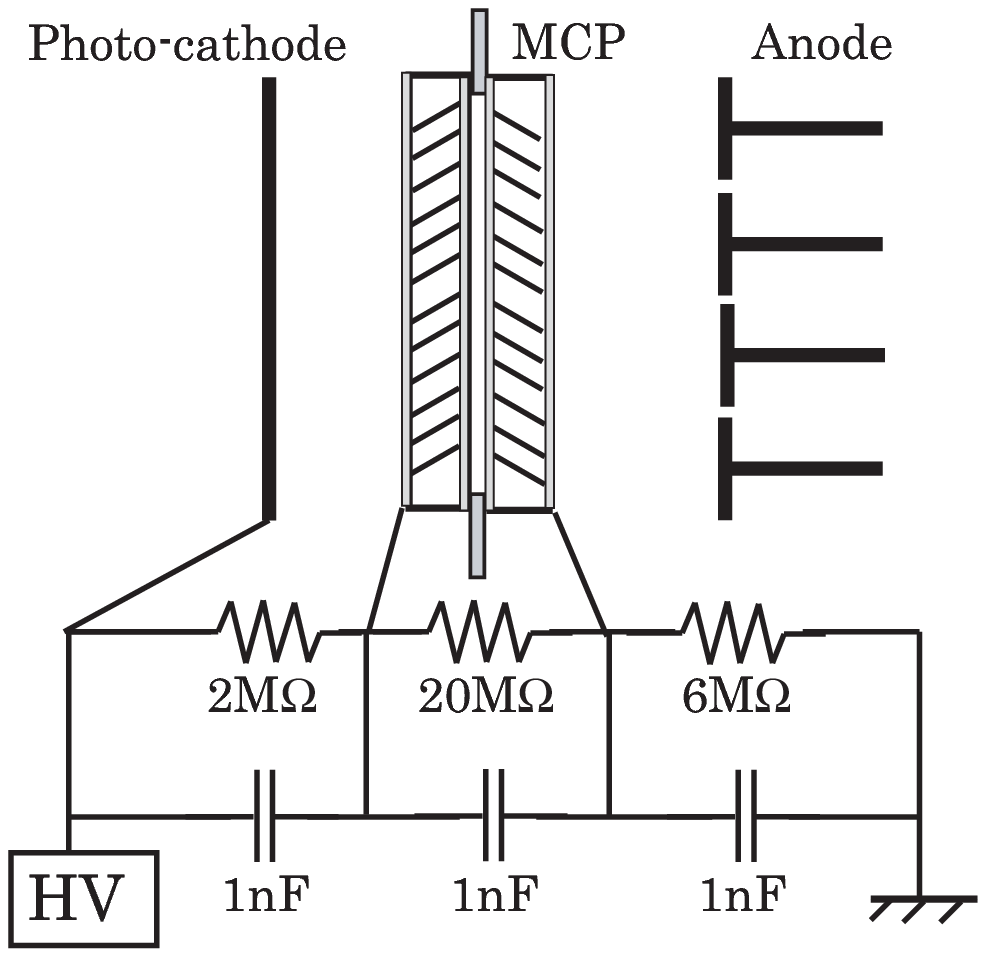}
}
\caption{\small Schematic drawing of the internal structure and the HV network 
of SL10-s.}
\label{F-2}
\end{figure}

A schematic of the internal structure and high-voltage (HV) supply network
of SL10-s is drawn in Fig.~\ref{F-2}. 
Both versions of SL10's have the same characteristics, 
as listed in Table~\ref{T-1}.
They are two-stage MCP's with a multi-alkali photocathode; the channel-hole
size and the thickness of an MCP-layer are 10 $\mu$m and 400 $\mu$m,
respectively. 
The anode plane has a size of 22$\times$22 mm$^2$, between which
there is a gap of 0.3 mm.

\begin{table}[h]
\caption{\small SL10 characteristics.}
\vspace*{2mm}
\label{T-1}
\begin{center}
\begin{tabular}{l|c}\hline
PMT size & 27.5 $\times$ 27.5 $\times$ 15.6 mm$^3$\\
Effective area & 22 $\times$ 22 mm$^2$ \\
Photocathode & Multi-alkali \\
Number of MCP layers~ & 2\\
MCP channel diameter~ & 10 $\mu$m\\
MCP aperture & 60 \%\\
MCP bias anngle & 13$^\circ$ \\
Photocathode$-$MCP gap & 2 mm \\
MCP$-$MCP gap & 0.03 mm \\
MCP$-$Anode gap & 1 mm \\
Al protection layer & No \\
Anode  & 4 $\times$ 1 linear array \\
Anode width & 5.3 mm\\ 
Anode gap & 0.3 mm \\
Supplied maximum HV & 3.5 kV \\
\hline
Quantum efficiency & 20 \% at 350 nm \\
Correction efficiency & $\sim 50$ \% \\
Gain & $2 \times 10^6$ at 3.5 kV \\
Transit time spread & 29 ps at 3.5 kV \\
Dark count & $< 10$ kHz/ch \\
\hline
\end{tabular}
\end{center}
\end{table}

\section{Performance of SL10} 

The performance of SL10, square-shaped and a multi-anode PMT, is measured
for single photons, and compared with those of a similar type of MCP-PMT's,
but with a round-shaped and a single anode (Hamamatsu Photonics K.K. (HPK), 
R3809U-50-11X and -25X, 
referred to as HPK6 (the channel-hole diameter of 6 ${\mu}$m) and 
HPK10 (10 ${\mu}$m) in \cite{MCPtiming}, respectively).

\subsection{Set-up}

A pico-sec light pulser (HPK, PLP-02-SLDH-041) is used to generate 
light of wavelength $\lambda=405 \pm 10$ nm 
with a pulse width of $34$ ps at a repetition rate of 1 kHz. 
Its intensity is attenuated down to a single-photon level by diffusers
and filters, and is then led to the SL10 surface by an optical fiber.  
This system is set inside a magnet in the case of a measurement of 
the SL10 response based on the B field strength, where the SL10 axis is placed 
so as to be parallel along the B field. 

The waveform of the output signals of the anode is recorded 
by a digital oscilloscope (Hewlett Packard, Infinium). 
To measure the charge and timing of the output pulse from individual anodes, 
it is fed into an attenuator (Agilent, 8495B:
attenuation = 0 - 70 dB, frequency $<$18 GHz) and then to an amplifier 
(HPK, C5594: gain= 36 dB, frequency = 50 kHz - 1.5 GHz). 
One of the amplifier outputs is fed into a CAMAC ADC (0.25 pC/count) module,
and the other output is sent into a CAMAC TDC (25 ps/count) module 
through a discriminator (Phillips Scientific, model-708) 
with a threshold voltage of 20 mV. 
It is $\sim$1/7 of the pulse height. 

A drawing of a similar setup is shown in Fig.~1 in \cite{MCPtiming}.

\subsection{Performance for single-photon detection}

Data presented in this subsection were obtained for the SL10-s, but there is 
no essential difference from that of the SL10-m.

\subsubsection{Raw signal and B field}

\begin{figure}[h]
\centerline{
\includegraphics[width=9cm]{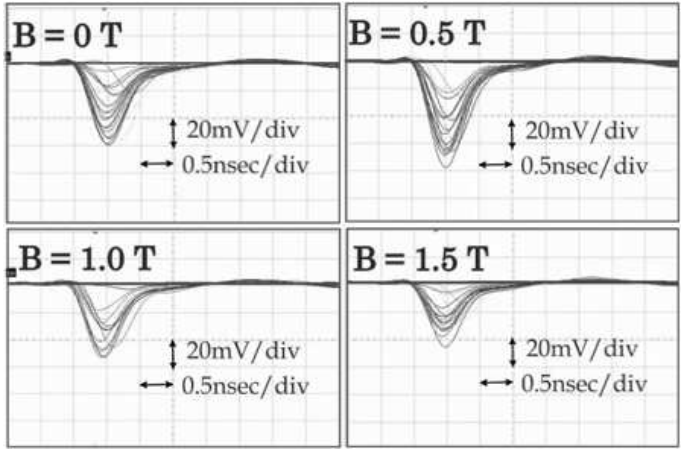}
}
\caption{\small Raw signals of the SL10-s anodes for single photons
under B = 0 $-$ 1.5 T, with HV = 3.5 kV.} 
\label{F-3}
\end{figure}

Figures~\ref{F-3} show the output signals for single photons under 
various B strengths. The pulse height is $\geq$30 mV and 
the rise time is $\simeq$500 ps at HV = 3.5 kV. 
These properties are consistent with those observed 
by the round-shaped PMT with a single anode, 
particularly HPK6 (see \cite{MCPtiming}). 

\subsubsection{Multiplication gain vs. B field}

\begin{figure}[h]
\centerline{
\includegraphics[width=8cm]{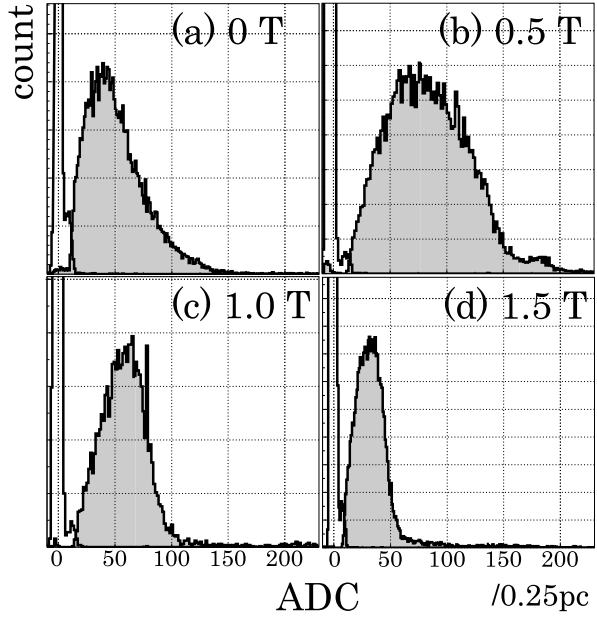}
}
\caption{\small Charge distributions for single photons under different 
B field strengths: (a) 0 T, (b) 0.5 T, (c) 1.0 T and (d) 1.5 T 
with HV = 3.5 kV.}
\label{F-4}
\end{figure}

\begin{figure}[h]
\centerline{
\includegraphics[width=8cm]{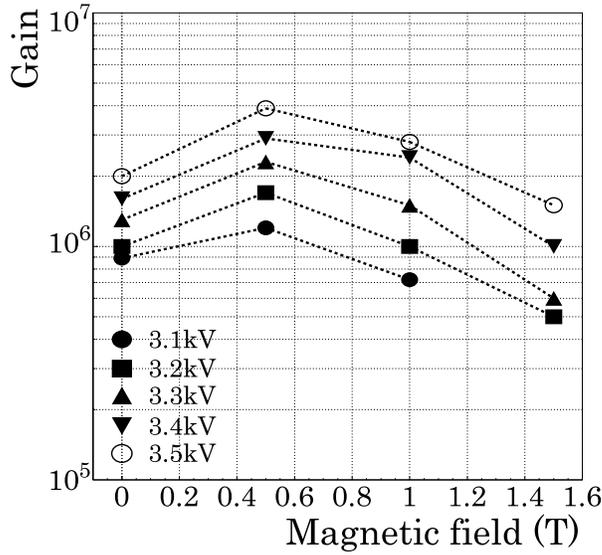}
}
\caption{\small $G$ vs B relations for single photons with different HV.} 
\label{F-4.5}
\end{figure}

Figure~\ref{F-4} shows the charge distributions of the output pulses 
under various B fields. 
A single photon peak is clearly seen around the 45-th, 70-th, 60-th 
and 40-th channel
at B=0, 0.5, 1.0 and 1.5 T, respectively.
From these output charges for single photons, the typical multiplication 
gain $G$ was evaluated, as indicated in Fig.~\ref{F-4.5}, 
with several different HV's from 3.1 kV to 3.5 kV with a 0.1 kV step:
The maximum $G$ of $4\times 10^6$ was attained at B = 0.5 T. 
The behavior of $G$ vs B is quite similar to that of HPK10, but 
its absolute magnitude is 5-10 times larger than that of HPK10. 
This difference is inferred to be due to a variation of individual 
MCP manufacturing batches.


\subsubsection{$G$ vs. photon rate} 

A huge rate of medium/high energy photons could be produced 
regarding the beams of the Super-KEKB accelerator;
their $e^+ e^-$ pair creation of 
the TOP quartz radiator then yielded many Cherenkov photons as backgrounds. 
At our TOP-counter configuration~\cite{Config}, 
the photon background rate at the Super-KEKB 
was estimated to be $\leq 3.4 \times 10^5$/cm$^2$/s
on the PMT surface~\cite{MCPlifetime}, by linearly extrapolating from 
the single rate of the present Belle TOF counter. 
The irradiation of a PMT by such a high-rate of photons could cause
a $G$ drop due to a finite recovery time of the effective HV field inside 
the channel-hole, apart from a shortening of the PMT lifetime.

\begin{figure}[h]
\centerline{
\includegraphics[width=8cm]{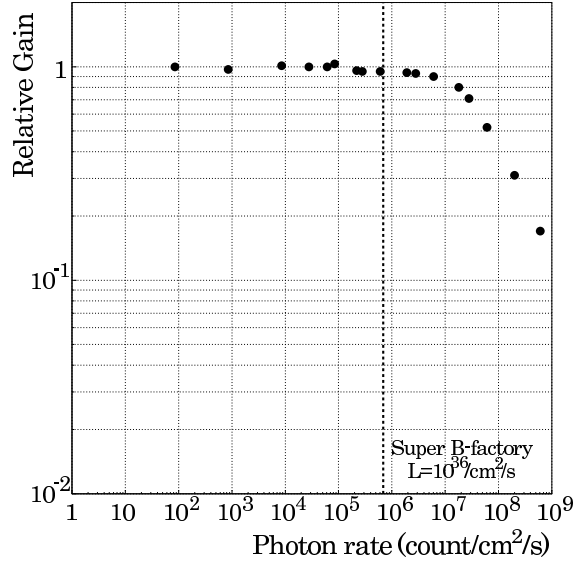}
}
\caption{\small Relative $G$ vs photon rate. 
Expected photon rate on the SL10 surface at Super-KEKB is indicated 
by the dashed vertical line.}
\label{F-5}
\end{figure}

The photon rate on the SL10 was varied by changing the light-pulser
frequency from 1 kHz up to 1 MHz, and the light intensity from 1 up to 600
photons/pulse. 
The resulting $G$ variation is shown in Fig.~\ref{F-5} as a function of
the photon rate. 
The SL10 stably functions up to a photon rate
of $\sim 1\times 10^7$/cm$^2$/s without significantly deteriorating its $G$. 
This rate corresponds to about 100-times higher than the expected rate at
the Super-KEKB factory. 

\subsubsection{TTS vs. B field}

\begin{figure}[h]
\centerline{
\includegraphics[height=5.8cm]{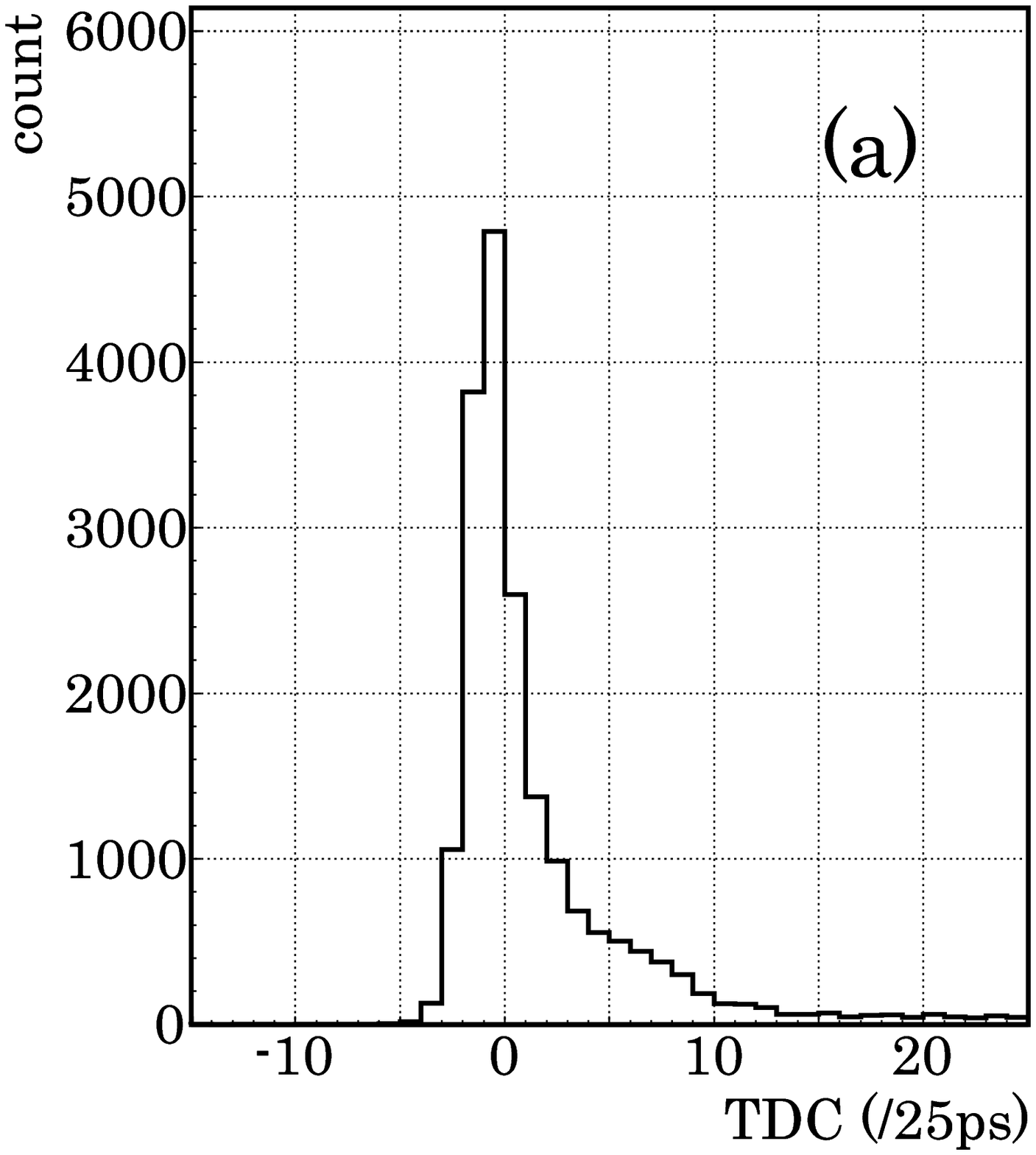}~~
\includegraphics[height=6cm]{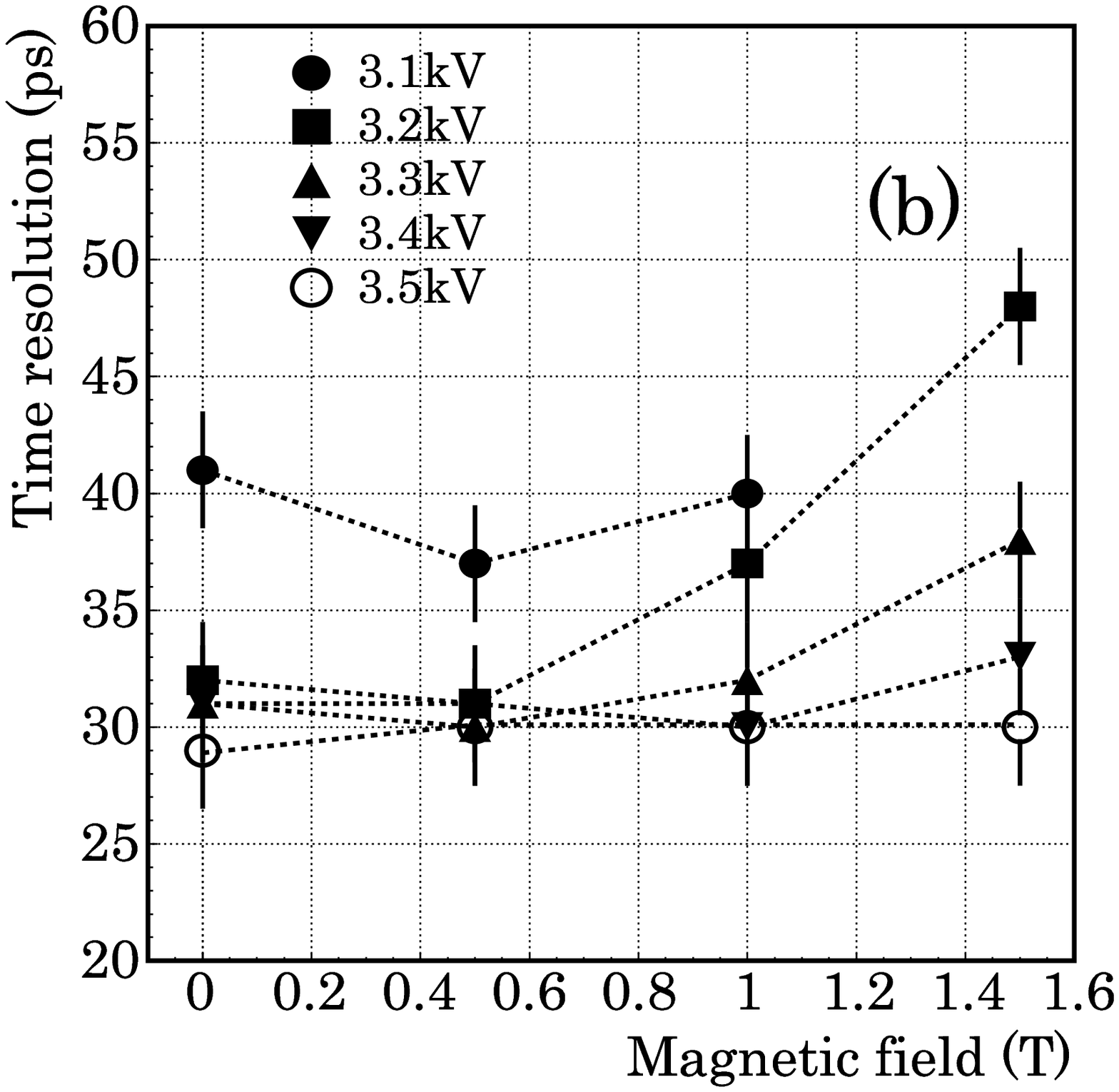}
}
\caption{\small (a) TDC distribution for single photons after
 the time-walk correction, and (b) $\sigma_{\rm TTS}$ as a function of B. }
\label{F-6}
\end{figure}

The time-walk corrected time response for single photons is plotted in 
Fig.~\ref{F-6}(a).
The dominant peak ($-125 {\rm ps} <$~TDC~$< 75 {\rm ps}$) yields 
the time resolution of $\sigma_{\rm TTS} = 29$ ps. 
The tail component is supposed to be due to the delayed photoelectrons
bouncing from the surface of the first MCP layer.  
Fig.~\ref{F-6}(b) shows the relation of $\sigma_{\rm TTS}$ vs. B; 
$\sigma_{\rm TTS} \simeq 30$ ps can be attained in the case of $G > 10^6$. 
This performance is similar to those of both HPK6 and HPK10
(see Fig.~7 in \cite{MCPtiming}). 

\section{Cross-talk and Its measure}\label{SEC4}

The cross-talk among the anodes is a general issue for multi-anode detectors;
moreover, it is substantial for a precise time measurement of 
the single photons, as reported in \cite{X-PMT}. 
In the case of the TOP-counter, $\sim$20 Cherenkov photons could be detected by 
15 pieces of the SL10's, each of which is equipped with 4 anodes, 
on a side edge of the quartz-bar, when a charged track passes through the bar. 
There are then appreciable probabilities that plural SL10's detect $\geq 2$
photons in an event. 
A primary single-photon would induce cross-talk on other anodes
in the same SL10, 
so that the pulse shape of the secondary arriving single-photon on the anode 
would be affected, and thus deteriorate the timing information.

We studied and then reduced the influence of the cross-talk by modifying 
the electrode structure as in the case of the SL10-s to SL10-m. 

No B-field is applied in the following studies.

\subsection{Phenomena}

\begin{figure}[h]
\centerline{
\includegraphics[width=8cm]{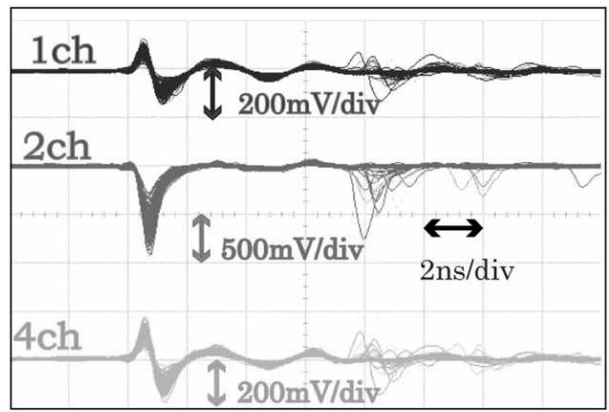}
}
\caption{\small Ch.2 exhibiting the signal pulses, and chs.1 and 4 
the cross-talk when ch.2 is irradiated by light pulses at 
$G\simeq 2\times 10^6$. }
\label{F-7}
\end{figure}

Fig.~\ref{F-7} shows the anode signals from the SL10-s in the case 
that an anode, channel 2, is irradiated by a sufficient number of photons,
$\sim$20, through a spot of 1 mm$^{\phi}$, to observe the cross-talk signals 
at any other channel. 
The cross-talk pulses show a differential shape of the signal pulse,
and its relative pulse height to the signal is $\sim$1/5 $-$ 1/3,
and does not depend 
on the distance among anodes that measure the signals and cross-talks. 
As is can be seen in Fig.~\ref{F-3}, the signal pulse height of 
single photons is
$\sim 40-50$ mV, so that the induced cross-talk height for single photons is 
below the threshold voltage of 20 mV. 
However, as demonstrated later in Fig.~\ref{F-10}, 
the base-line fluctuation caused 
by the cross-talk would make the time measurement ineffectual for a consecutive 
photon on a channel other than ch.2. 

Second signals can be seen at $\simeq 7$ ns after the primary signals
in Fig.~\ref{F-7}. 
It is caused by feedback $H^+$ ions, yielded in the multiplication process of 
the micro-channel plate; the second pulses also induce cross-talk 
at other anodes.

\subsection{Origin}

We extensively studied the cross-talk on a multi-anode 
16-channel linear array vacuum PMT~\cite{X-PMT}, and here 
just also found the same phenomena for the SL10.

\begin{figure}[h]
\centerline{
\includegraphics[width=7cm]{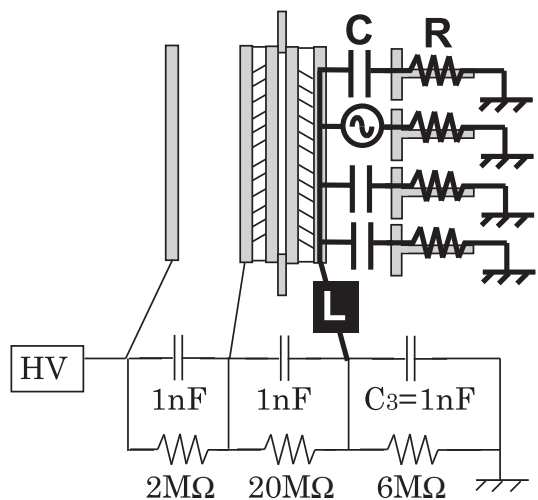}
}
\caption{\small Schematic drawing of an electric equivalent circuit from 
a cross-talk point of view on SL10-s.}
\label{F-8}
\end{figure}

The cross-talk in this case might be induced through, especially, the last 
electrode, facing the anodes, of the second MCP-layer.  
An internal structure of SL10-s, and its readout and HV circuits are 
illustrated in Fig.~\ref{F-8}.
The anode, $22\times 5.3$ mm$^2$, and the last electrode of the MCP 
separated by a 1 mm gap, would form a capacitance of $C \sim$1 pF; 
a short HV line from the HV bleeder network 
to the last electrode would yield an inductance of $L \sim$50 nH;
also, individual anodes are terminated with a resistance of $R = 50$ $\Omega$. 

\subsection{Measure}

\begin{figure}[h]
\centerline{
\includegraphics[width=8cm]{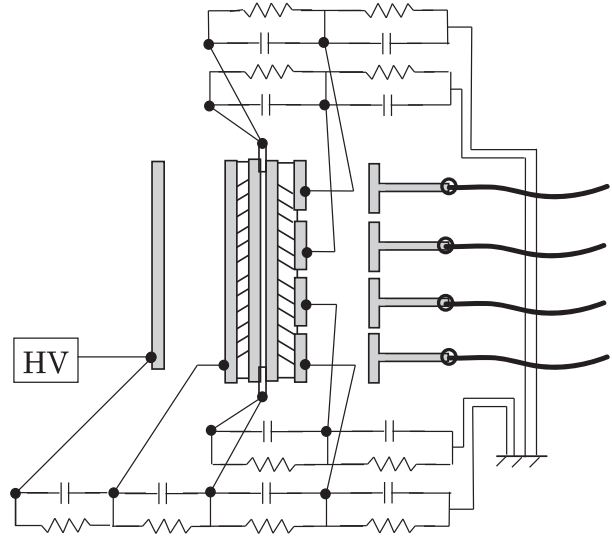}
}
\caption{\small Structure and HV network of the internally segmented SL10, SL10-m. }
\label{F-9}
\end{figure}

We thought out that one way to reduce the cross-talk by disconnecting this
RLC loop circuit is to make a segmentation of the MCP electrode facing 
to 4 anodes, and to prepare 4 separate HV networks, 
as illustrated in Fig.~\ref{F-9}. 
Since the MCP is a high-resistance material, 20 M$\Omega$ per plate, 
the impedances between the segmented electrodes are quite large, electrically 
well disconnected. 

\subsection{Performance of the segmented SL10, SL10-m}

\subsubsection{Signals}

\begin{figure}[h]
\centerline{
\includegraphics[width=8cm]{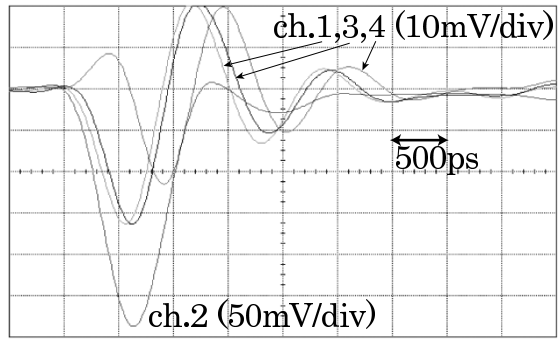}
}
\caption{\small Signal (ch.2) and cross-talk (chs.1, 3 and 4) 
of the SL10-m.}
\label{F-10}
\end{figure}

Anode signals of the SL10-m can be seen in Fig.~\ref{F-10}, where the setup
condition is the same in the case of Fig.~\ref{F-7}, except that 
the number of photons irradiated is $\sim$10, rather than $\sim$20. 
The pulse-height ratio between the cross-talk and the signal is now $\sim$10. 
It has been improved by a factor of $\sim$2, compared to that of SL10-s.  

\subsubsection{$\sigma_{\rm TTS}$ under cross-talk}

\begin{figure}[h]
\centerline{
\includegraphics[width=9cm]{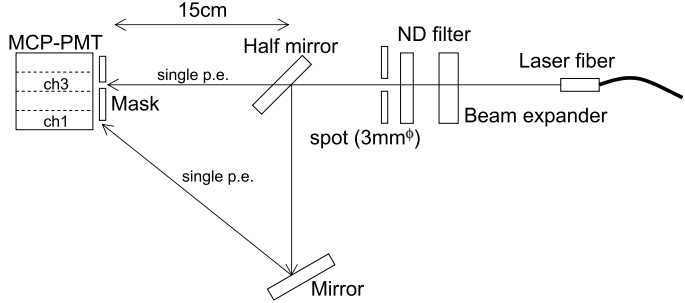}
}
\caption{\small Setup to measure the cross-talk effect on SL10's.}
\label{F-11}
\end{figure}

We prepared the setup shown in Fig.~\ref{F-11} to measure the time resolution 
in a case that two photons hit different anodes of the SL10-s and 
SL10-m within an adjacent narrow time period, 
supposing Cherenkov ring imaging photons in the TOP-counter. 
The light beams from the pulser are divided into two by a half mirror.
The primary light is detected by ch.3 through a mask, and 
the divided and then delayed light is by ch.1 as shown in Fig.~\ref{F-11}. 
The delay time, $t_{\rm d}$, is adjusted to be within a range of 
up to $t_{\rm d}\sim$3 ns, 
by changing the distance of the second mirror to the SL10. 
A light pulse is attenuated down to single-photon level by beam expanders
and filters, upstream of the half mirror.  

\begin{figure}[h]
\centerline{
\includegraphics[width=9cm]{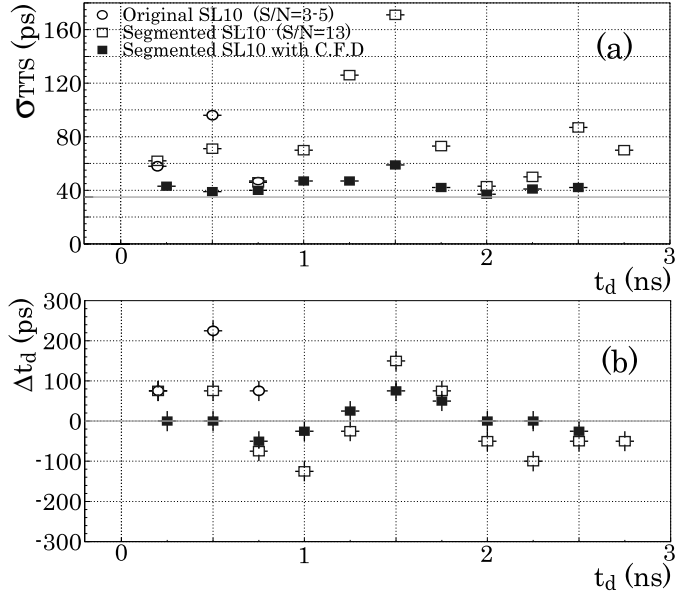}
}
\caption{\small (a) $\sigma_{\rm TTS}$ and (b) difference between 
the observed time and the delayed time for the delayed photons at ch.1. 
Data with open circles and open squares are those obtained by SL10-s and SL10-m,
respectively, using a voltage threshold type of discriminator; 
the closed squares indicate data by the SL10-m, but with constant fraction 
discrimination. } 
\label{F-12}
\end{figure}

The observed time resolution, $\sigma_{\rm TTS}$, for the delayed photons
is plotted in Fig.~\ref{F-12} as a function of $t_{\rm d}$ 
by open circles for SL10-s and open squares for SL10-m, 
using a discriminator of the voltage threshold type
(Phillips Scientific, model-708, threshold = 20 mV). 
For SL10-s, which provides $\sigma_{\rm TTS}\simeq 30$ ps in the case of 
single photons with no cross-talk 
(see, Fig.\ref{F-6}), 
there is no data point at $t_{\rm d} \geq 1$ ns because 
the discriminator hits the cross-talk pulse before the delayed signal arrives.

For SL10-m, 
the cross-talk pulse is sufficiently suppressed so that only the delayed
signal is detected by the discriminator at any $t_{\rm d}$ range.
However, $\sigma_{\rm TTS}$ strongly depends on $t_{\rm d}$.
For $t_{\rm d}\sim 1.5$ ns, around which the cross-talk effect from the primary 
signal is the largest, $\sigma_{\rm TTS}$ becomes $120 - 170$ ps, 
while in the other $t_{\rm d}$ region,
where the fluctuation of the base line is rather small, 
$\sigma_{\rm TTS} = 50 - 80$ ps. 
Even so, the latter resolution is worse than the $\sigma_{\rm TTS}\simeq 35$ ps 
attained at the incidence of sole single photons.

\subsubsection{Further measure with a constant fractional discriminator}

Although SL10-m successfully reduces the cross-talk, 
it does not recover the original time resolution of 
$\sigma_{\rm TTS}\simeq 35$ ps for single photons. 
The cross-talk effect also appears as a shift of the observed absolute time of 
the delayed photons, as plotted in Fig.~\ref{F-12}(b). 
The difference, $\Delta t_{\rm d}$, between the observed time and 
$t_{\rm d}$ behaves as expected from the cross-talk pulse 
(see, Fig.~\ref{F-10}).
When the base line shifts to the positive side in voltage, 
$\Delta t_{\rm d}$ moves to a positive delay, and vise-versa. 
This fact indicates that the signal pulse is not broken away by the cross-talk, 
rather it keeps its intrinsic signal shape upon a tiny cross-talk disturbance. 
 
We replaced a leading-edge type of discriminator with a constant fractional 
discriminator (CFD, Kaizu-works KN381), 
since the latter provides an output pulse at zero-crossing timing 
of a differentiated input pulse so that it would not be strongly affected by 
the cross-talk, as the former would be.
The resulting $\sigma_{\rm TTS}$ and $\Delta t_{\rm d}$ with the cross-talk
are plotted by closed squares with error bars in Fig.~\ref{F-12}. 
The time resolution largely improved to $\sigma_{\rm TTS} = 40 - 60$ ps 
with no large variation in $t_{\rm d}$. 
The timing shift has also been reduced by a factor of 2: 
$\Delta t_{\rm d}=\pm 50$ ps.
However, because its maximum shift is about the same size 
as the time resolution
it is not a trivial matter, unless it is corrected by referring to the timing
of the primary signal.

\section{Summary and Discussions}

We have developed a square-shaped 4-anode MCP-PMT, SL10, to measure the precise 
timing of single photons.  
It provides a performance that satisfies the requirement on our TOP-counter: 
$G\sim 10^6$, $\sigma_{\rm TTS} \leq 35$ ps for single photons under a 1.5 T 
magnetic field strength, and durable up to $\sim 1\times 10^7$ /cm$^2$/s of 
a high photon rate.  

In order to suppress the cross-talk effect on the timing measurement
in the case of plural photon injection, we have taken two measures. 
One way is to subdivide the electrode of the last MCP-layer into 4 pieces,
facing the anodes: It successfully reduces the cross-talk size to one half. 
The second way is to use a constant fractional discriminator, instead of 
the leading-edge type one. The time resolution is largely recovered up to 
$\sigma_{\rm TTS}=40-60$ ps, while not yet restoring the original resolution 
of $\sigma_{\rm TTS}\simeq 35$ ps.

{\ }\\
\noindent
{\bf Acknowledgments}\\

This work is supported by a Grant-in-Aid for Science Research on Priority Area 
(Mass Origin and Supersymmetry Physics, 18071003) from the Ministry of 
Education, Culture, Sports, Science and Technology of Japan, and
by a Grant-in-Aid for Creative Scientific Research (18GS0206)
from the Japan Society for the Promotion of Science.


\begin{thebibliography}{99}
\bibitem{Config} 
A segment of TOP counters will comprize two different lengths of 
the quartz radiators: 2,000/1,000 mm$^L\times$ 400 mm$^W\times$ 20 mm$^t$, 
each of which equips 15 pieces of SL10's at an end. 
Such 18 segments will form a barrel with 
r=1,200 mm from the beam-colliding point.
\end{thebibliography}
\end{document}